\documentclass[twocolumn,twoside]{revtex4}
\usepackage{graphicx}
\usepackage{fancyhdr}
\usepackage{epsfig}
\usepackage{amsmath}
\def\DAF{DA$\Phi$NE} 
\def\ifm#1{\relax\ifmmode#1\else$#1$\fi}

\newcommand{\eV}{{e\kern-.07em V}}

\newcommand{\MeV}{{\rm \,M\eV}}

\newcommand{\GeV}{{\rm \,G\eV}}

\makeatletter
\newdimen\z@ \z@=0pt 
\newskip\z@skip \z@skip=0pt plus0pt minus0pt
\def\m@th{\mathsurround=\z@}
\def\ialign{\everycr{}\tabskip\z@skip\halign} 
\def\eqalign#1{\null\,\vcenter{\openup\jot\m@th
  \ialign{\strut\hfil$\displaystyle{##}$&$\displaystyle{{}##}$\hfil
    \crcr#1\crcr}}\,}
\makeatother
\newcommand{\BR}{B\kern -0.1em R}
\newcommand{\dd}{{\rm d}\,}

\begin{document}

\title{KLOE extraction of $a_\mu^{\pi\pi}$ in the mass range
$[0.35,0.95]\mbox{ GeV}^2$}

\author{The KLOE Collaboration\\
F.~Ambrosino, A.~Antonelli, M.~Antonelli, F.~Archilli,
C.~Bacci, P.~Beltrame, G.~Bencivenni,
S.~Bertolucci, C.~Bini, C.~Bloise, S.~Bocchetta, V.~Bocci, F.~Bossi, P.~Branchini,
R.~Caloi, P.~Campana, G.~Capon, T.~Capussela, F.~Ceradini, S.~Chi, G.~Chiefari, P.~Ciambrone, 
E.~De~Lucia, A.~De~Santis, P.~De~Simone, G.~De~Zorzi, A.~Denig, A.~Di~Domenico, C.~Di~Donato,
S.~Di~Falco, B.~Di~Micco, A.~Doria, M.~Dreucci, G.~Felici,
A.~Ferrari, M.~L.~Ferrer, G.~Finocchiaro, S.~Fiore, C.~Forti, P.~Franzini, C.~Gatti,
P.~Gauzzi, S.~Giovannella, E.~Gorini, E.~Graziani, M.~Incagli, W.~Kluge, V.~Kulikov,
F.~Lacava, G.~Lanfranchi, J.~Lee-Franzini, D.~Leone,
M.~Martini, P.~Massarotti, W.~Mei, L.~Me\-o\-la, S.~Mi\-scet\-ti, M.~Moulson, S.~M\"uller, F.~Murtas,
M.~Napolitano, F.~Nguyen, 
M.~Palutan, E.~Pasqualucci, A.~Passeri, V.~Patera, F.~Perfetto, M.~Primavera,
P.~Santangelo, G.~Saracino, B.~Sciascia,
A.~Sciubba, F.~Scuri, I.~Sfiligoi,
T.~Spadaro, 
M.~Testa, L.~Tortora, 
P.~Valente, B.~Valeriani, G.~Venanzoni, R.~Versaci,
G.~Xu}

\begin{abstract}
The KLOE Experiment at the $\phi$ factory \DAF\, has measured
the cross section $\sigma(e^+e^-\to\pi^+\pi^-\gamma)$
using two different selection schemes:
requiring the photon emission at small polar angle 
and detecting the photon at large polar angle in the calorimeter.
Using a theoretical radiator function we extract the pion form factor and 
obtain the $\pi\pi$ contribution to the anomalous magnetic moment of the muon. 
Results presented here come from the analysis of 240 pb$^{-1}$ collected in 2002,
with improved systematic 
uncertainty with respect to the published KLOE analysis. We also include an
update of the previous analysis.
\end{abstract}

\maketitle

\thispagestyle{fancy}

\section{Introduction}
\label{sec:Intro}
The muon magnetic anomaly is one of the
best known quantities in Particle Physics.
Recent measurements
performed at Brookhaven, reach an accuracy of 0.54 ppm~\cite{Bennett:2006fi}.
Theorists derive a number that differs from the experimental value
by 3.2 standard deviations~\cite{Jegerlehner:2007xe}.
The main source of uncertainty of the theoretical
estimate stems from the hadronic contributions,
not calculable within the perturbative regime of QCD.
The 
hadronic contribution at the
lowest order, $a_\mu^{hlo}$, is obtained
from a dispersion integral of hadronic
cross section measurements.
Incidentally, the annihilation reaction
with the final state $\pi^+\pi^-$ 
accounts for $\sim70\%$ of $a_\mu^{hlo}$
and $\sim60\%$ of the uncertainty.

\section{{\bf \DAF}\ and KLOE}
\label{sec:DaEKlo}
DA$\Phi$NE is an $e^+ e^-$ collider running 
at $\sqrt{s}\simeq M_\phi$,
the $\phi$ meson mass, which has
provided 
an integrated luminosity of about 2.5 fb$^{-1}$
to the KLOE experiment up to year 2006.
In addition, about 250 pb$^{-1}$ of data have been collected
at $\sqrt{s}\simeq 1$ GeV, in 2006.

Present results are based on 240 pb$^{-1}$
of data taken in 2002, sufficient to
reach a statistical relative error
less than 0.2\% in the $\pi\pi$ contribution
on the muon magnetic anomaly, $a_\mu^{\pi\pi}$.

The KLOE detector consists of a drift chamber~\cite{Adinolfi:2002uk} with
excellent momentum resolution ($\sigma_p/p\sim 0.4\%$
for tracks with polar angle larger than $45^\circ$)
and an electromagnetic calorimeter~\cite{Adinolfi:2002zx} with good energy
($\sigma_E/E\sim 5.7\%/\sqrt{E~[\mathrm{GeV}]}$)
and precise time ($\sigma_t\sim 54~\mathrm{ps}/\sqrt{E~[\mathrm{GeV}]}\oplus
50~\mathrm{ps}$) resolution.

\section{Measurement of the $\sigma_{\pi\pi}$ cross section}
\label{sec:MeTo}
At DA$\Phi$NE, we measure the differential spectrum 
of the $\pi^+\pi^-$ invariant mass, $M_{\pi\pi}$, from
Initial State Radiation (ISR) events,
$e^+ e^-\to\pi^+\pi^-\gamma$, and extract
the total cross section $\sigma_{\pi\pi}\equiv\sigma_{e^+ e^-\to\pi^+\pi^-}$
using the following formula~\cite{Binner:1999bt}:
\begin{equation}
M_{\pi\pi}^2~ \frac{\dd\sigma_{\pi\pi\gamma}}
{\dd M_{\pi\pi}^2} = \sigma_{\pi\pi}
(M_{\pi\pi}^2)~ H(M_{\pi\pi}^2)~,
\label{eq:1}
\end{equation}
where $H$ is the radiator function.
This formula neglects
Final State Radiation (FSR) terms.

\subsection{Small angle $\gamma$ selection, SA}
The cross section for ISR photons has a pronounced
divergence in the forward angle (relative to the beam
direction), then overwhelmingly dominates over
FSR photon production.
Thus, the fiducial volume
of the SA event selection scheme, both for the
published 2001~\cite{Aloisio:2004bu}
and 2002 data 
are based on the following selection criteria:
\begin{itemize}
\item[a)] two tracks with opposite charge within the
polar angle range $50^\circ<\theta<130^\circ$, this helps
in obtaining good reconstructed tracks;
\item[b)] small angle photon, $\theta_\gamma<15^\circ\, (\theta_\gamma>165^\circ)$,
the photon is not explicitly detected and its direction
is reconstructed from the track momenta,
$\vec{p}_\gamma=-(\vec{p}_{\pi^+} +\vec{p}_{\pi^-})$,
this enhances the probability that it is an ISR photon.
\end{itemize}
This fiducial volume implies the following advantages:
\begin{itemize}
\item FSR at the Leading Order is reduced to the $0.3\%$ level;
\item the contamination from the resonant process $e^+e^-\to
  \phi\to\pi^+\pi^-\pi^0$ -- where at least one of photons coming
from the $\pi^0$ is lost -- is reduced to the level of $\sim5\%$.
\end{itemize}
Discrimination of $\pi^+\pi^-\gamma$
from $e^+ e^-\to e^+ e^-\gamma$ events
is done 
via particle identification~\cite{knote:192}
based on the time of flight,
on the shape and the energy
of the clusters associated to the
tracks. In particular, electrons deposit
most of their energy in the
first planes of the calorimeter while
minimum ionizing muons and pions release uniformly the
same energy in each plane.
An event is selected if at least one
of the two tracks has not being identified
as an electron. This criterion
results in a rejection power of 97\%
for $e^+ e^-\gamma$ events, while retaining a selection
efficiency of $\sim 100\%$ for $\pi^+\pi^-\gamma$ events.

Contaminations from the processes
$e^+e^-\to\mu^+\mu^-\gamma$ and
$\phi\to\pi^+\pi^-\pi^0$ are rejected
by cuts on the track mass variable, $m_{trk}$,
defined by the four-momentum conservation --
assuming a final state consisting
of two particles with the same mass and one photon --
\begin{figure}
\begin{center}
\includegraphics[width=15.6pc,height=13.3pc]{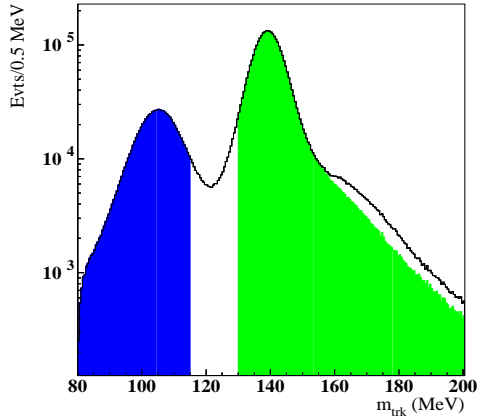}
\vglue-0.4cm
\caption{Discrimination of the $\mu^+\mu^-\gamma$ (filled dark histogram) from the
$\pi^+\pi^-\gamma$ (filled light histogram) events, after the selection
discussed in the text. The residual
contamination of $\pi^+\pi^-\pi^0$ events is evident at high $m_{trk}$ values.}
\label{fig:1}       
\end{center}
\end{figure}
and on the missing mass, $m_{miss}=
\sqrt{E^2_{X}-|\vec{P}_{X}|^2}$, defined assuming
the process is $e^+ e^-\to\pi^+\pi^-X$.
Fig.~\ref{fig:1} shows the separation between
$\pi^+\pi^-\gamma$ and $\mu^+\mu^-\gamma$ events, achieved
selecting different $m_{trk}$ regions around the
mass peaks. 

Nevertheless, the small $\gamma$ angle selection does not
allow us to cover the threshold region,
where the photon takes the maximum
available momentum and the pions recoil
the opposite direction with minimum opening angle.

\subsection{Large angle $\gamma$ selection, LA}
The threshold region becomes accessible as the photon
is emitted into the same solid angle of the pion tracks. 
Thus, a different KLOE analysis is done
requiring the detection of at least
one photon of energy larger than 50 MeV and
$50^\circ<\theta_\gamma<130^\circ$ in the calorimeter.
This allows the closure of the kinematics with the
following variables:
\begin{itemize}
\item the angle between the photon direction and the missing momentum,
$\vec{p}_{miss}=-(\vec{p}_{\pi^+} +\vec{p}_{\pi^-})$;
\item the $\chi^2$ of a kinematic fit implementing four-momentum conservation
and $\pi^0$ mass as constraints.
\end{itemize}
These requirements are applied to reject the $\pi^+\pi^-\pi^0$
contamination -- much larger than in the SA requirement.
Moreover, this LA selection is sensitive to larger FSR effects,
including interference from the resonant~\cite{Ambrosino:2005wk} decays $\phi\to f_0(980)\gamma$,
with $f_0(980)\to\pi^+\pi^-$ and $\phi\to\rho^\pm\pi^\mp$,
with $\rho^\pm\to\pi^\pm\gamma$. The phases and couplings
of these interference effects have to be obtained from Monte Carlo
using phenomenological models~\cite{Czyz:2004nq,Pancheri:2006cp}.

\subsection{Improvements with respect to the published
analysis} 
The analysis of data taken since 2002
benefits from cleaner and more stable running conditions
of \DAF\, 
resulting in less machine background. 
In particular the following changes are applied with respect to
the data taken in 2001:
\begin{itemize}
\item an additional trigger level was implemented at the end of 2001
to eliminate a 30\% loss due
to pions penetrating up to the outer calorimeter plane and being misidentified
as cosmic rays events (from 2002 on, this inefficiency has decreased
down to 0.2\% on $\pi^+\pi^-\gamma$ events);
\item the offline background filter, which 
contributed the largest experimental systematic uncertainty to the published
work~\cite{Aloisio:2004bu}, has been improved and includes now a downscale algorithm providing an 
unbiased control sample. This greatly facilitates
the evaluation of the filter efficiency which increased from 95\% to 98.5\%,
with negligible systematic uncertainty.
\end{itemize}
In addition to the aforementioned items, 
the knowledge of the detector response and
of the KLOE simulation program 
has been improved.

\section{Luminosity}
\label{sec:Lumi}
The absolute normalization of the data sample is measured using
large angle Bhabha scattering events, $55^\circ<\theta<125^\circ$, with a cross
section $\sigma\simeq430$ nb. The integrated luminosity, $\mathcal{L}$, is
provided~\cite{Ambrosino:2006te} dividing the observed number
of these events by the
effective cross section evaluated by the Monte Carlo generator
\texttt{Babayaga}~\cite{Carloni Calame:2000pz},
including QED radiative corrections
with the parton shower algorithm, inserted in the code simulating
the KLOE detector. Recently, an updated version of the generator,
\texttt{Babayaga@NLO}~\cite{Balossini:2006wc}, has been released.
\begin{table}[htbp]
\begin{center}
{\small
 \renewcommand{\arraystretch}{1.6}
 \setlength{\tabcolsep}{0.9mm}
\begin{tabular}{|l|c|}
\hline
\multicolumn{2}{|c|}{systematic errors on $\mathcal{L}$ (\%)}\\
\hline
theory, $\sigma_{th}$ & 0.10 \\
\hline
acceptance & 0.25 \\
background & 0.08 \\
$e^\pm$ reconstruction & 0.13 \\
energy calibration & 0.10 \\
knowledge of $\sqrt{s}$ & 0.10 \\
\hline
total: $\sigma_{th}\oplus\sigma_{exp}$ & 0.3 \\
\hline
\end{tabular}
}
\end{center}
\caption{\label{tab:1}
Relative systematic error on
the luminosity measurement: experimental contributions
are listed in detail and added in quadrature to yield the
relative experimental error, $\sigma_{exp}$.}
\end{table}
The new predicted cross section
decreased by 0.7\% and the theoretical systematic uncertainty
improved from 0.5\% to 0.1\%. The experimental
uncertainty is summarized in Table~\ref{tab:1}
and is dominated by differences in the angular acceptance
between data and MC.

\section{Extraction of $a_\mu^{\pi\pi}$}
The differential cross section is obtained from the observed spectrum,
$N_{obs}$, after subtracting
the re\-si\-dual background events, $N_{bkg}$, and correcting for
the selection efficiency, $\varepsilon_{sel}(M_{\pi\pi}^2)$,
and the luminosity:
\begin{equation}
\frac{\dd\sigma_{\pi\pi\gamma}}
{\dd M_{\pi\pi}^2} = \frac{N_{obs}-N_{bkg}}
{\Delta M_{\pi\pi}^2}\, \frac{1}{\varepsilon_{sel}(M_{\pi\pi}^2)~ \mathcal{L}}~ ,
\label{eq:2}
\end{equation}
where our mass resolution allows us to have bins of
width $\Delta M_{\pi\pi}^2=0.01\GeV^2$.
The above formula is used in both small and large $\gamma$ angle analyses.

The residual background content is found fitting the
$m_{trk}$ spectrum of the selected data sample with a superposition
of Monte Carlo distributions describing the signal and
background sources. The only free parameters of these fits
are the relative weights of signal and backgrounds in data,
computed as a function of $M_{\pi\pi}$.

The radiator function used to obtain $\sigma_{\pi\pi}$
as in eq.(\ref{eq:1}) is provided by the code
\texttt{Phokhara}~\cite{Rodrigo:2001kf}, setting the pion form factor $F_\pi(M_{\pi\pi})=1$.
In addition, $\sigma_{\pi\pi}$ is corrected for the running of the fine
structure constant~\cite{Jegerlehner:2006ju} (vacuum polarization) and for shifting from $M_{\pi\pi}$
to the virtual photon mass, $M_{\gamma^*}$, for those events
with both an initial and a final photon. The \texttt{Phokhara}~\cite{Czyz:2005as}
version also having this contribution is used for the
acceptance correction and for all efficiencies using
the kinematics from Monte Carlo.

This corrected and
FSR inclusive cross section, $\sigma_{\pi\pi(\gamma)}^{bare}$,
is used to determine $a_\mu^{\pi\pi}$:
\begin{equation}
\label{eq:3}
a_\mu^{\pi\pi} = 
\frac{1}{4\pi^3}\int_{s_{low}}^{s_{up}}\dd s~
\sigma_{\pi\pi(\gamma)}^{bare}(s)\,K(s)~ ,
\end{equation}
where the lower and upper bounds of the spectrum measured
with the small $\gamma$ angle analysis are respectively
$s_{low}=0.35\GeV^2$ and $s_{up}=0.95\GeV^2$.

\subsection{Comparison with the published result}
\label{sec:Comp2001}
Table~\ref{tab:2} shows
the list of relative systematic uncertainties
in the evaluation of $a_\mu^{\pi\pi}$ in the mass range
[0.35,0.95] GeV$^2$, for the published analysis of 2001
data and for the \emph{preliminary} analysis of 2002 data.
The comparison is performed for the same small $\gamma$ angle
selection.
\begin{table}[htbp]
\begin{center}
{\small
 \renewcommand{\arraystretch}{1.6}
 \setlength{\tabcolsep}{0.9mm}
\begin{tabular}{|l|c|c|c|}
\hline
\multicolumn{4}{|c|}{systematic errors on $a_\mu^{\pi\pi}$ (\%)}\\
\hline
~ & 2001 analysis & ~ & 2002 analysis\\
\hline
offline filter & 0.6 & & negligible\\
background & 0.3 & & 0.3\\
kinematic cuts & 0.2 & & 0.2\\
$\pi$/e ID & 0.1 & & 0.3\\ 
vertex & 0.3 & & 0.5\\
tracking & 0.3 & & 0.4\\
trigger & 0.3 & & 0.2\\
acceptance & 0.3 & & 0.1\\
FSR corrections & 0.3 & & 0.3\\
luminosity & 0.6 & & 0.3\\
$H$ function eq.(\ref{eq:1}) & 0.5 & & 0.5\\
vacuum polarization & 0.2 & & negligible\\
\hline
total & 1.3 & & 1.1 \\
\hline
\end{tabular}
}
\end{center}
\caption{\label{tab:2}
Comparison of relative systematic errors on
the extraction of $a_\mu^{\pi\pi}$ in the mass range
[0.35,0.95] GeV$^2$ between the analysis of 2001
data and the \emph{preliminary} analysis of 2002 data,
using the same small $\gamma$ angle selection.}
\end{table}

In revisiting the published analysis we found a bias in the evaluation
of the trigger correction that affects mostly the low $M_{\pi\pi}$
\begin{figure}
\begin{center}
\includegraphics[width=19.2pc,height=15.5pc]{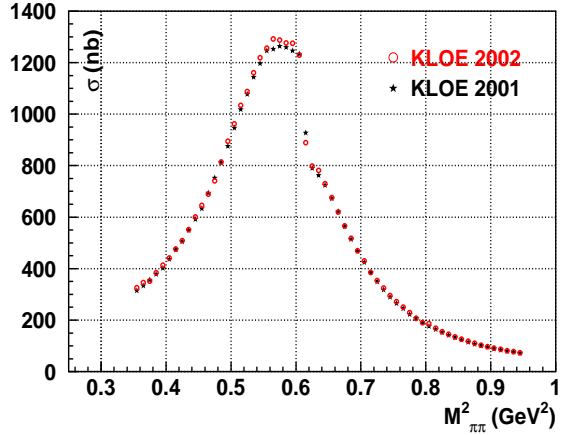}
\vglue-0.4cm
\caption{Comparison between $\sigma_{\pi\pi}$ measured in 2001 (dark star),
updated for the trigger correction and for the new
theoretical Bhabha cross section, and that measured in 2002 (light circle).}
\label{fig:2}       
\end{center}
\end{figure}
region. Correcting for this effect
and normalizing to the new Bhabha cross section
we updated the 2001 spectrum to compare with the new one from
the 2002 analysis. Fig.~\ref{fig:2} shows this comparison
\begin{table}[htbp]
\begin{center}
{\small
 \renewcommand{\arraystretch}{1.6}
 \setlength{\tabcolsep}{0.9mm}
\begin{tabular}{|l|c|}
\hline
\multicolumn{2}{|c|}{$a_\mu^{\pi\pi}([0.35,0.95]\GeV^2)\times10^{10}$}\\
\hline
2001 published & $388.7~\pm~0.8_{stat}~\pm~4.9_{sys}$\\
\hline
2001 updated & $384.4~\pm~0.8_{stat}~\pm~4.9_{sys}$\\
\hline
2002 preliminary & $386.3~\pm~0.6_{stat}~\pm~3.9_{sys}$\\
\hline
\end{tabular}
}
\end{center}
\caption{\label{tab:3}
Comparison among $a_\mu^{\pi\pi}$ values evaluated with
the small $\gamma$ angle selection.}
\end{table}
in terms of $\sigma_{\pi\pi}$.

The net shift is below
one stadard deviation
on $a_\mu^{\pi\pi}$. Table~\ref{tab:3} shows
these results, which are in excellent
agreement.

\subsection{Comparison between small and large $\gamma$ angle
selections}
An excellent cross check of the $a_\mu^{\pi\pi}$
evaluation at small angle
\begin{figure}
\begin{center}
\includegraphics[width=19.2pc,height=15.5pc]{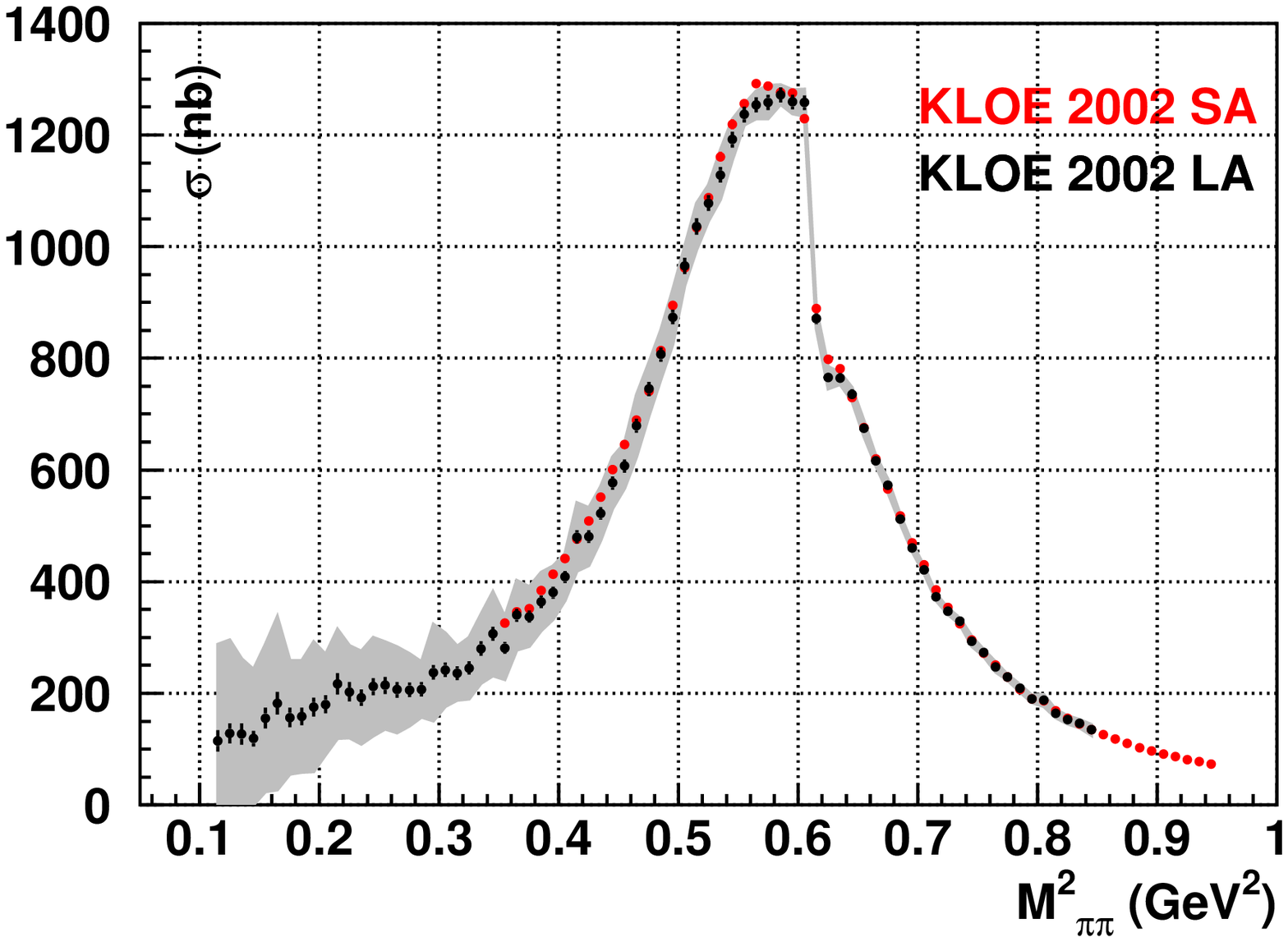}
\includegraphics[width=16.5pc,height=12.5pc]{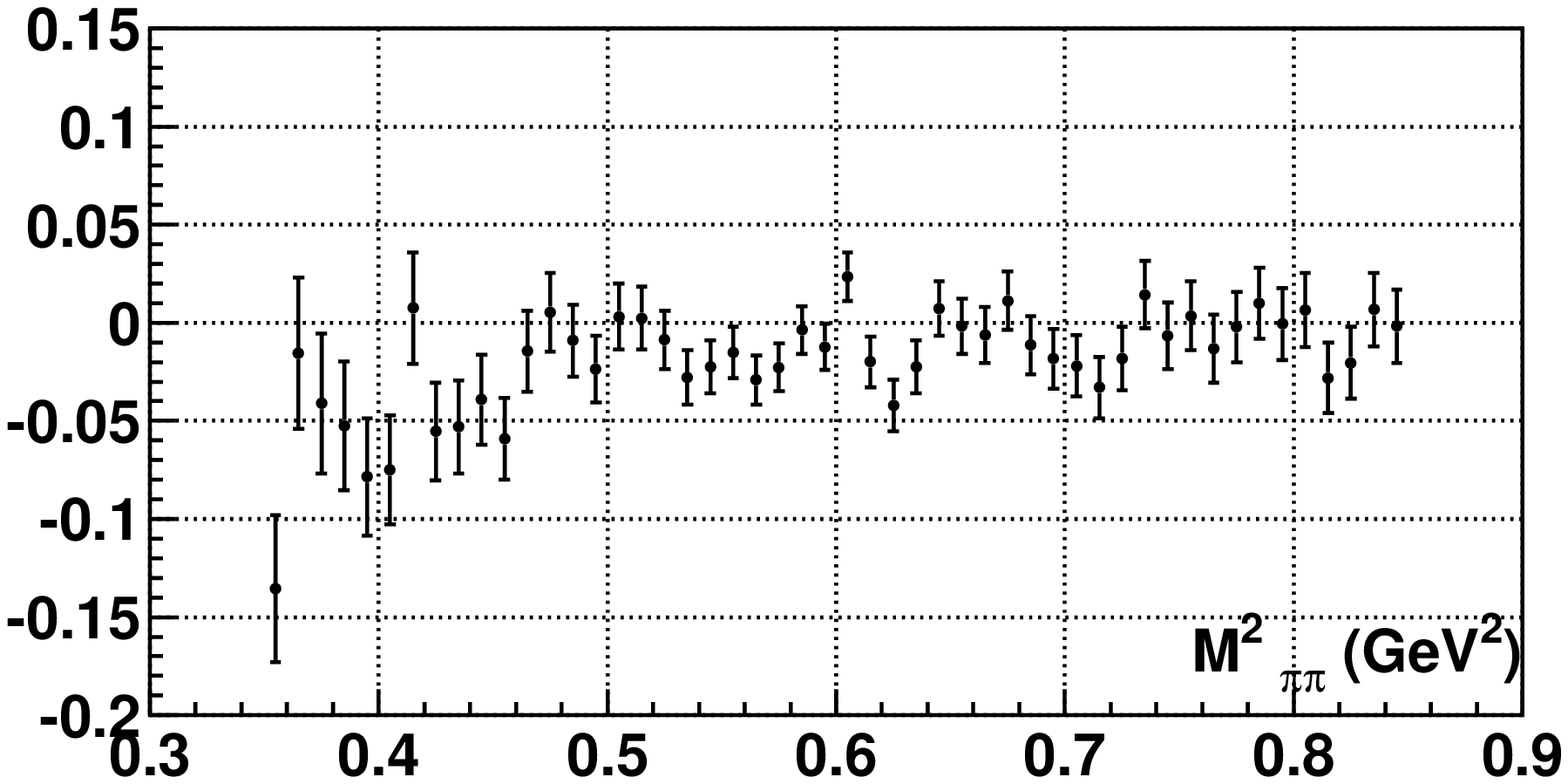}
\vglue-0.3cm
\caption{Top: comparison between the small (SA) and large (LA) $\gamma$ angle
selections in $\sigma_{\pi\pi}$. In the latter case, also
the systematic uncertainty is accounted for, as a band. Bottom: relative difference
between the two cross sections shown in the upper panel.}
\label{fig:3}       
\end{center}
\end{figure}
is from the measurement of $\sigma_{\pi\pi}$
with photon emitted at large angle. Since it is an independent
event selection scheme, that allows us also to test the 
knowledge of FSR contributions. Fig.~\ref{fig:3} shows
the comparison between the two spectra. In the large angle
case, the main source of uncertainty is the unknown
interference between the FSR process and the resonant decays
$\phi\to\pi^+\pi^-\gamma$ -- mentioned before --
limiting the accuracy at low and high $M_{\pi\pi}$ va\-lues, such
\begin{table}[htbp]
\begin{center}
{\small
 \renewcommand{\arraystretch}{1.6}
 \setlength{\tabcolsep}{0.9mm}
\begin{tabular}{|l|c|}
\hline
\multicolumn{2}{|c|}{$a_\mu^{\pi\pi}([0.5,0.85]\GeV^2)\times10^{10}$}\\
\hline
2002 small angle & $255.4~\pm~0.4_{stat}~\pm~2.5_{sys}$\\
\hline
2002 large angle & $252.5~\pm~0.6_{stat}~\pm~5.1_{sys}$\\
\hline
\end{tabular}
}
\end{center}
\caption{\label{tab:4}
Comparison between $a_\mu^{\pi\pi}$ values evaluated with
the small and large $\gamma$ angle selections.}
\end{table}
that our comparison is limited to the range
[0.5,0.85] GeV$^2$. In that range we evaluated and
compared $a_\mu^{\pi\pi}$. Table~\ref{tab:4} shows
the good agreement for $a_\mu^{\pi\pi}$ between the two independent
measurements. The main source of systematic
uncertainty in the large $\gamma$ angle selection
is from $f_0(980)$ background subtraction.

\subsection{Preliminary comparison between KLOE and VEPP-2M
experiments}
We also made a comparison with the most recent
$a_\mu^{\pi\pi}$ evaluations released by the CMD-2~\cite{Akhmetshin:2006bx} and
SND~\cite{Achasov:2006vp}
experiments, in the mass range $M_{\pi\pi}\in[630,958]\MeV$.
\begin{table}[htbp]
\begin{center}
{\small
 \renewcommand{\arraystretch}{1.6}
 \setlength{\tabcolsep}{0.9mm}
\begin{tabular}{|l|c|}
\hline
\multicolumn{2}{|c|}{$a_\mu^{\pi\pi}(M_{\pi\pi}\in[630,958]\MeV)\times10^{10}$}\\
\hline
CMD-2~\cite{Akhmetshin:2006bx} & $361.5~\pm~1.7_{stat}~\pm~2.9_{sys}$\\
\hline
SND~\cite{Achasov:2006vp} & $361.0~\pm~2.0_{stat}~\pm~4.7_{sys}$\\
\hline
KLOE preliminary & $355.5~\pm~0.5_{stat}~\pm~3.6_{sys}$\\
\hline
\end{tabular}
}
\end{center}
\caption{\label{tab:5}
The $a_\mu^{\pi\pi}$ value presented here compared with
other recent evaluations.}
\end{table}
Table~\ref{tab:5} shows this comparison, where the published
CMD-2 and SND values agree with the new 
KLOE result within one standard deviation.

While the CMD-2 and SND dispersion integrals are performed
with the trapezoid rule, the KLOE value is done extrapolating
our bins to match the [630,958] MeV mass range, and summing directly
the bin contents of the $\dd\sigma_{\pi\pi\gamma}$ differential
spectrum -- weighed for the kernel function.

\section{Conclusions and outlook}
We obtained the $\pi\pi$ contribution to $a_\mu$ in the mass range
$M_{\pi\pi}^2\in[0.35,0.95]\GeV^2$ integrating
the differential cross section for the ISR events
$e^+ e^-\to\pi^+\pi^-\gamma$, measured with the KLOE detector:
\begin{enumerate}
\item the preliminary result from 2002 data agrees with the updated
  result from the published KLOE result, based on the small $\gamma$ angle
  analysis of 2001 data;
\item the preliminary 2002 results both from the large and small
  $\gamma$ angle analyses agree in a region where FSR effects play a minor
  role, $M_{\pi\pi}^2\in[0.5,0.85]\GeV^2$;
\item the small angle 2002 result is also in agreement within
  one standard deviation with the recent SND
  and CMD-2 values
  in the mass range $M_{\pi\pi}\in[630,958]\MeV$.
\end{enumerate}
Further work is going on towards final results:
\begin{itemize}
\item refining the small angle analysis, \textit{i.e.} by including
  the effects due to unfolding the detector resolution;
\item improving the knowledge of the FSR interference effects for the
  large angle analysis, using KLOE $f_0(980)$
  measurements~\cite{Ambrosino:2005wk,Ambrosino:2006hb};
\item measuring the pion form factor directly from the ratio bin-by-bin
  of $\pi^+\pi^-\gamma$ to $\mu^+\mu^-\gamma$ spectra~\cite{Muller:2006bk};
\item extracting the pion form factor with a large angle selection from
  data taken at $\sqrt{s}=1$ GeV, off the $\phi$ resonance.
\end{itemize}

\end{document}